\documentclass{emulateapj}

\usepackage{lscape}

\begin{document}
\journalinfo{Accepted for publication in \textit{The Astrophysical Journal Letters}}
\title{Leo V: Spectroscopy of a Distant and Disturbed Satellite\footnote{Observations reported here were obtained at the MMT Observatory, a facility operated jointly by the University of Arizona and the Smithsonian Institution.}}
\shorttitle{Spectroscopy of Leo V}
\author{M.G. Walker\altaffilmark{1}, V. Belokurov\altaffilmark{1},
  N.W. Evans\altaffilmark{1}, M.J. Irwin\altaffilmark{1},
  M. Mateo\altaffilmark{2}, E.W. Olszewski\altaffilmark{3}, G. Gilmore\altaffilmark{1}}
\email{walker@ast.cam.ac.uk}
\altaffiltext{1}{Institute of Astronomy, University of Cambridge, UK}
\altaffiltext{2}{Department of Astronomy, University of Michigan, Ann Arbor}
\altaffiltext{3}{Steward Observatory, The University of Arizona, Tucson, AZ}

\begin{abstract} 
  We present a spectroscopic study of Leo V, a recently discovered
  satellite of the Milky Way (MW).  From stellar spectra obtained with
  the MMT/Hectochelle spectrograph we identify seven likely members of
  Leo V.  Five cluster near the Leo V center ($R < 3 \arcmin$) and
  have velocity dispersion $2.4_{-1.4}^{+2.4}$ km s$^{-1}$.  The other
  two likely members lie near each other but far from the center ($R
  \sim 13\arcmin \sim 700$ pc) and inflate the global velocity
  dispersion to $3.7_{-1.4}^{+2.3}$ km s$^{-1}$.  Assuming the five
  central members are bound, we obtain a dynamical mass of $M=
  3.3_{-2.5}^{+9.1}\times 10^5M_{\odot}$ ($M/L_V=
  75_{-58}^{+230}[M/L_V]_{\odot}$).  From the stacked spectrum of the
  five central members we estimate a mean metallicity of [Fe/H]$=
  -2.0\pm 0.2$ dex.  Thus with respect to dwarf spheroidals of similar
  luminosity, Leo V is slightly less massive and slightly more
  metal-rich.  Since we resolve the central velocity dispersion only
  marginally, we do not rule out the possibility that Leo V is a
  diffuse star cluster devoid of dark matter.  The wide separation of
  its two outer members implies Leo V is losing mass; however, its
  large distance ($D\sim 180$ kpc) is difficult to reconcile with MW
  tidal stripping unless the orbit is very radial.
\end{abstract}

\section{Introduction}

The recent discoveries of thirteen ultra-faint Milky Way (MW)
satellites (e.g., \citealt{willman05b,zucker06b,belokurov07}) have
reshaped the census of Local Group galaxies and increased our ability
to test cosmological models using objects available in our own
neighborhood.  The new satellites, discovered primarily with data from
the Sloan Digital Sky Survey (SDSS; \citealt{york00}), extend the
galaxy luminosity function by three orders of magnitude, to $\sim
10^3$L$_{\odot}$ \citep{koposov08}.  Spectroscopic surveys (e.g.,
\citealt{martin07}, \citealt[SG07 hereafter]{simon07}) of their few
red giants reveal that these systems extend scaling relationships
according to which the least luminous dwarf galaxies are the most
dominated by dark matter and most metal-poor (e.g.,
\citealt{mateo98,kirby08}).

Leo V, the most recent addition to the ensemble of MW satellites, was
originally detected as a modest ($\sim 4\sigma$) overdensity of stars
in SDSS data \citep[``Paper I'' hereafter]{belokurov08}.  Slightly
deeper photometry from the Isaac Newton Telescope (INT) reveals a blue
horizontal branch (BHB) and red clump in addition to the red giant
branch (RGB; Figure 3 of Paper I).  The apparent magnitude of its BHB
implies Leo V has $M_V\sim -4.3$ and lies at a distance of $\sim 180$
kpc, placing it at a (deprojected) distance of just $\sim 20$ kpc from
another SDSS dSph, Leo IV \citep{belokurov07}.  As the systemic
line-of-sight velocities of Leos IV and V differ by just $\sim 40$ km
s$^{-1}$, the two objects may represent the smallest companions among
the known MW satellites.

With this letter we present an initial spectroscopic study of
individual stars in Leo V.  Despite heavy contamination by the
Galactic foreground, we are able to identify seven likely members of
Leo V.  We use the stacked spectrum of these members to estimate Leo
V's mean metallicity, and we use their spatial and velocity
distributions to investigate Leo V's dynamical state and plausible
dark matter content.

\section{Observations,  Data \& Membership}

On 2008 May 28 and 30 we obtained medium-resolution spectra of 158 Leo
V targets using the Hectochelle spectrograph at the MMT Observatory.
We maximized the chance of observing Leo V members by choosing targets
based on proximity to the center of Leo V and to the locus of its RGB
(Figure 3 of Paper I).  We extracted and calibrated spectra following
the procedure described in detail by \citet{mateo08}.  The Hectochelle
spectra span the wavelength range $5150-5300$ \AA, in which the most
prominent absorption feature is the Mg-I/Mg-b triplet (MgT).  We
measure the line-of-sight velocity of each star by cross-correlating
its spectrum against co-added Hectochelle spectra of radial velocity
standards.  We also measure a composite magnesium index, $\Sigma$Mg,
from the weighted sum of pseudo-equivalent widths for each individual
line of the MgT (see \citealt{walker07a}).

We follow SG07 in modelling the velocity error as the sum of random
and systematic components: the error in the $i^{th}$ measurement is
$\sigma_i=\sqrt{\sigma_{{\rm ran},i}^2+\sigma_{\rm sys}^2}$.  We
determine the random errors, which correlate with S/N, using a
bootstrap method.  For the $i^{th}$ spectrum, we generate 1000
artificial spectra by adding randomly generated Poisson deviates to
the counts at each pixel.  After measuring the velocity for each
artificial spectrum, we equate $\sigma_{{\rm ran},i}^2$ with the
variance of the artificial distribution.  We then use repeat
Hectochelle observations of 98 stars (including 11 Leo V targets) to
determine the systematic error common to all measurements.  The value
$\sigma_{{\rm sys}}=0.35$ km s$^{-1}$ best satisfies our requirement
that the distribution of $(V_1-V_2)(\sigma_{{\rm
    ran},1}^2+\sigma_{{\rm ran},2}^2+2\sigma_{{\rm sys}}^2)^{-1/2}$
resemble a Gaussian with unit variance.  By the same procedure, we
estimate the systematic error of $\Sigma$Mg to be $0.091$ \AA.  Among
likely Leo V members in our sample, the mean (median) velocity error
is $\pm 2.2$ km s$^{-1}$ ($\pm 1.6$ km s$^{-1}$).

Table \ref{tab:table} presents spectroscopic data for the 11 stars
that have velocity in the range $162 < V/$ (km s$^{-1})<180$, around
the systemic mean reported in Paper I.  The complete data set is
available as an electronic table.  

\begin{figure}
  \epsscale{1.0}
\begin{center}
\includegraphics[width=0.45\textwidth]{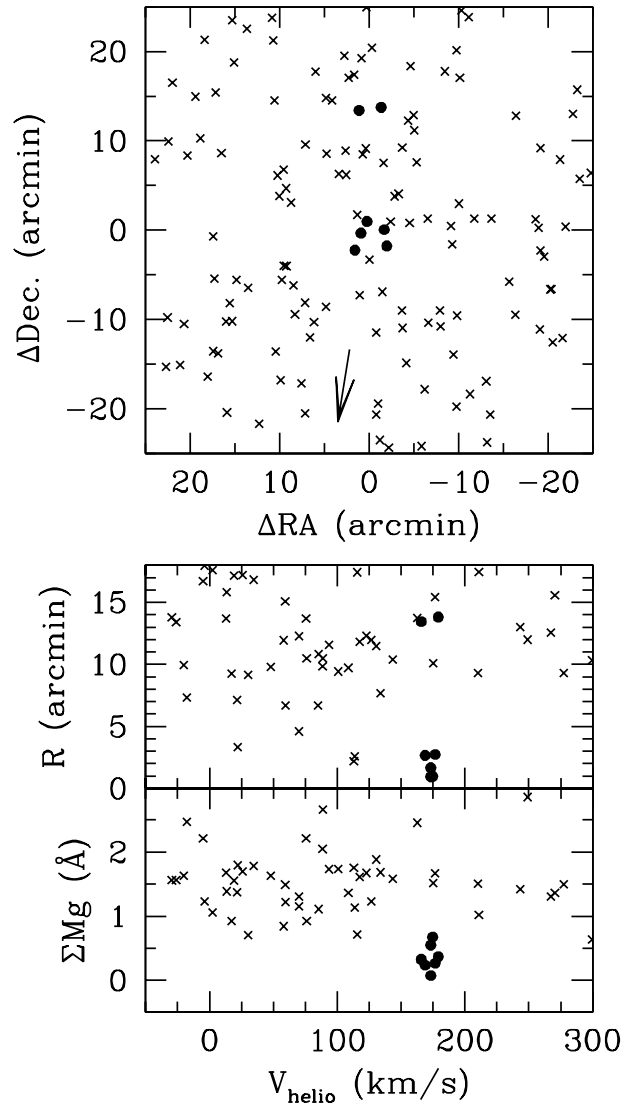}
\end{center}
  \caption{\scriptsize MMT/Hectochelle specotroscopic data for red
    giant candidates in Leo V.  \textit{Top:} Sky positions of
    measured stars.  Solid circles (crosses) indicate likely members
    (nonmembers), for which the EM algorithm returns $P_{M} > 0.5$
    ($P_{M}<0.5$).  An arrow points toward the Leo IV satellite, which
    has central coordinates $(27, -165)$.  \textit{Middle:} Projected
    distance from the Leo V center vs.\ line-of-sight velocity.
    \textit{Bottom:} Magnesium index vs.\ velocity.  The two members
    at large radius fall in the middle of Leo V's $\Sigma$Mg
    distribution.}
  \label{fig:members_leo5}
\end{figure}

Leo V's low luminosity implies that most of the stars overlapping its
red giant branch are late-type dwarfs in the Milky Way foreground.  To
the eye, the Leo V population appears in Figure \ref{fig:members_leo5}
as a clump with narrow velocity distribution at small radius and weak
$\Sigma$Mg.  We quantify this separation using an
expectation-maximization (EM) algorithm similar to those devised for determining membership in open clusters (e.g., \citealt{sanders71}).  Our algorithm (see \citealt{walker09b} for details)
iteratively evaluates $V$ and $\Sigma$Mg distributions and assigns to
each star a probability, $\hat{P}_{M}$, of belonging to the member
population.  The algorithm converges after identifying seven stars as
likely members of Leo V.

Three stars (L5-31, L5-60 and L5-116 in Table \ref{tab:table}) that
would have been considered members after applying a conventional
$3\sigma$ velocity threshold are identified as foreground
($\hat{P}_{M} <0.01$), based on strong $\Sigma$Mg and large distance
from the Leo V center.  Two other stars (L5-52 and L5-57) at large
radius ($R\sim 13\arcmin$) are likely members ($\hat{P}_{M} =0.7\pm 0.3$
and $\hat{P}_{M}=0.8\pm 0.3$, respectively, where the reported error is derived from Monte Carlo tests in which we repeat the EM algorithm after resampling $V$ and $\Sigma$Mg from the sets of artificial spectra) based on their weak $\Sigma$Mg
values, which fall almost exactly on the mean value for Leo V.  We
find no members between $3\arcmin < R < 13\arcmin$.  Given Leo V's
measured half-light radius of $r_{\rm h}=0.8\arcmin \pm 0.1\arcmin$ (Paper
I), the presence of two members beyond this gap is unexpected unless
Leo V's morphology is highly distorted (see Section
\ref{sec:discussion}).
 
 \renewcommand{\arraystretch}{0.6}
 \begin{deluxetable*}{lrcrrrrrrrrrrrrrrrrrrr}
 \tabletypesize{\scriptsize}
 \tablewidth{0pc}
 \tablecaption{\scriptsize Hectochelle Spectroscopy of Leo V (abridged: complete table is available in electronic version) }
 \tablehead{\colhead{Target}&\colhead{$\alpha_{2000}$}&
 \colhead{$\delta_{2000}$}&\colhead{$R$}&\colhead{$r$}&\colhead{$g-r$}&
 \colhead{$V_{helio}$}&\colhead{$\Sigma$Mg}&\colhead{$\hat{P}_{M}$}&\\
 \colhead{}&\colhead{(hh:mm:ss)}&\colhead{(dd:mm:ss)}&\colhead{(arcmin)}&
 \colhead{(mag)}&\colhead{(mag)}&\colhead{(km s$^{-1}$)}&\colhead{(\AA)}&
 \colhead{}\\
 }
 \startdata
L5-001&11:31:13.21&$+02$:12:51.6&  1.0& 20.43    &  0.64    &$  173.4\pm   3.8$&$ 0.07\pm  0.38$&$1.000\pm 0.000$\\
L5-002&11:31:10.59&$+02$:14:09.5&  1.0& 19.85    &  0.78    &$  174.8\pm   0.9$&$ 0.68\pm  0.16$&$1.000\pm 0.000$\\
L5-004&11:31:02.88&$+02$:13:14.7&  1.7& 19.53    &  0.71    &$  173.2\pm   1.5$&$ 0.55\pm  0.16$&$0.996\pm 0.003$\\
L5-007&11:31:01.66&$+02$:11:25.2&  2.7& 19.67    &  0.74    &$  168.8\pm   1.6$&$ 0.24\pm  0.32$&$0.994\pm 0.005$\\
L5-008&11:31:15.91&$+02$:10:57.6&  2.7& 19.53    &  0.73    &$  176.8\pm   2.1$&$ 0.27\pm  0.22$&$0.996\pm 0.002$\\
L5-031&11:31:46.64&$+02$:09:10.9& 10.1& 20.80    &  0.70    &$  175.2\pm   1.8$&$ 1.52\pm  0.19$&$0.000\pm 0.000$\\
L5-052&11:31:14.11&$+02$:26:36.4& 13.4& 20.90    &  0.50    &$  165.6\pm   2.4$&$ 0.39\pm  0.25$&$0.735\pm 0.353$\\
&&&&&&$  166.0\pm   2.6$&$ 0.23\pm  0.20$&\\
L5-055&11:30:14.90&$+02$:14:30.1& 13.7& 20.55    &  0.59    &$  162.8\pm   1.6$&$ 2.46\pm  0.32$&$0.000\pm 0.000$\\
L5-057&11:31:04.15&$+02$:26:56.5& 13.8& 21.31    &  0.42    &$  179.2\pm   3.7$&$ 0.37\pm  0.21$&$0.788\pm 0.304$\\
L5-060&11:30:53.03&$+01$:58:20.2& 15.4& 19.94    &  0.77    &$  176.7\pm   3.2$&$ 1.67\pm  0.52$&$0.000\pm 0.290$\\
L5-116&11:32:22.01&$+01$:56:49.3& 24.4& 21.00    &  1.00    &$  174.6\pm   2.0$&$ 0.90\pm  0.29$&$0.000\pm 0.000$\\
 \enddata
 \label{tab:table}
 \end{deluxetable*}

\section{Velocity Dispersion and Dynamical Mass}
\label{sec:vdisp}

Assuming Leo V has a Gaussian velocity distribution with mean $\langle
V \rangle$ and variance $\sigma_{V_0}^2$, the data have 2D likelihood
$L(\langle V \rangle,\sigma_{V_0})\propto\prod_{i=1}^N
[(\sigma_{V_0}^2+\sigma_{V_i}^2)^{-1/2}\exp [(V_i-\langle
V\rangle)^2/(2[\sigma_{V_i}^2+\sigma_{V_0}^2])]^{M_i}$, where exponent
$M=1$ for member stars and $M=0$ for contaminants.  The values $M_i$
are unknown, but we can use the membership probabilities as
weights in evaluating the expected log-likelihood given by\footnote{Our results are insensitive to our implicit assumption that the velocity error distributions are Gaussian.  If we instead use the exact error distributions we obtain from the sets of artificial spectra, we obtain the same constraints on the velocity dispersion.}
\begin{eqnarray}
  E(\ln L)=-\frac{1}{2}\displaystyle\sum_{i=1}^N \hat{P}_{M_i}\ln (\sigma_{V_0}^2+\sigma_{V_i}^2)\hspace{1in}\nonumber\\
  -\frac{1}{2}\displaystyle\sum_{i=1}^N \hat{P}_{M_i} \biggl [\frac{(V_i-\langle V \rangle)^2}{\sigma_{V_0}^2+\sigma_{V_i}^2} \biggr ]+\mathrm{const}.\hspace{0.15in}
  \label{eq:loglike}
\end{eqnarray}
Because we are interested primarily in $\sigma_{V_0}^2$, we follow
\citet{kleyna04} in considering the one-dimensional likelihood given
by $L_{1D}(\sigma_{V_0}^2)=\int_{-\infty}^{+\infty}L(\langle
V\rangle,\sigma_{V_0}^2)d\langle V\rangle$.  Again treating the
probabilities $\hat{P}_M$ as weights, we evaluate the 1D expected
likelihood
\begin{eqnarray}
  E(\ln L_{1D})=-\frac{1}{2}\ln a-\frac{1}{2}\biggl (c-\frac{b^2}{a}\biggr ) \hspace{1in}\nonumber\\
  -\frac{1}{2}\displaystyle\sum_{i=1}^N\hat{P}_{M_i}\ln (\sigma_{V_i}^2+\sigma_{V_0}^2)+\mathrm{const},\hspace{0.15in}
  \label{eq:loglike1d}
\end{eqnarray}
where $a=\sum_{i=1}^N \hat{P}_{M_i}/(\sigma_{V_i}^2+\sigma_{V_0}^2)$,
$b=\sum_{i=1}^N \hat{P}_{M_i}V_i/(\sigma_{V_i}^2+\sigma_{V_0}^2)$, and
$c=\sum_{i=1}^N \hat{P}_{M_i}V_i^2/(\sigma_{V_i}^2+\sigma_{V_0}^2)$.
We obtain error bounds by evaluating the area beneath the curve given
by $\hat{L}_{1D}=\exp[E(\ln L_{1D})]$.  Because of the bimodal
distribution of $R$ among the likely Leo V members, we measure two
sets of distribution parameters.  Hereafter, ``global'' refers to the
entire sample (7 members), while ``central'' pertains to a sample
restricted to stars with $R<5\arcmin$ ($5$ members).

\begin{figure}
\begin{center}
\includegraphics[width=0.45\textwidth]{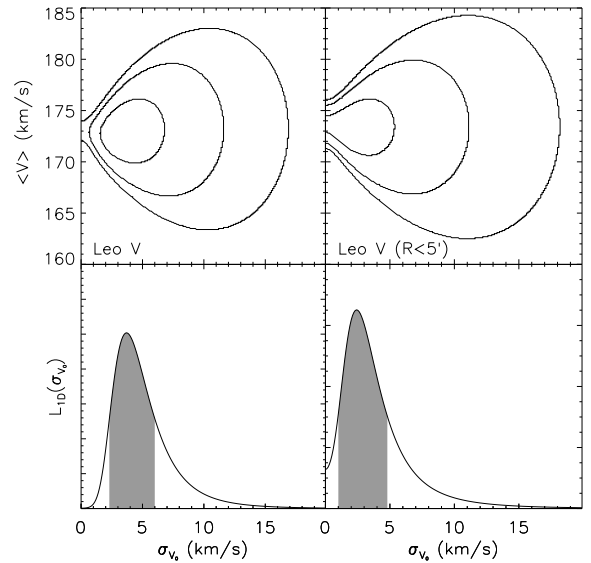}
\end{center}
  \caption{\scriptsize Mean velocity and velocity dispersion of Leo V,
    from the entire kinematic sample ($\sim 7$ members; left panels)
    and from stars within $5\arcmin$ of the Leo V center ($\sim 5$
    members; right panels).  Contours in upper panels enclose $68\%$,
    $95\%$ and $99\%$ of the volume underneath the likelihood surface
    described by Equation \ref{eq:loglike}.  Lower panels give the 1D
    likelihood of the velocity dispersion, with the shaded region
    marking the central $68\%$ of the area under the curve.}
  \label{fig:leo5_vdisp}
\end{figure}

Figure \ref{fig:leo5_vdisp} (top) displays contours in the $(\langle V
\rangle,\sigma_{V_0})$ plane that enclose $68\%$, $95\%$, and $99\%$
of the volume underneath the surface given by $\hat{L}=\exp[E(\ln
L)]$.  While the mean velocity and global velocity dispersion are well
constrained, we resolve the central velocity dispersion only
marginally.  This result holds also when we consider the 1D likelihood
of the velocity dispersion (lower panels in Figure
\ref{fig:leo5_vdisp}).  We measure a global velocity dispersion of
$\sigma_{V_0}=3.7_{-1.4 (-2.3)}^{+2.3 (+6.6)}$ km s$^{-1}$ and a
central velocity dispersion of $\sigma_{V_0}=2.4_{-1.4 (-2.4)}^{+2.4
  (+7.0)}$ km s$^{-1}$, where errors indicate $68\%$ ($95\%$)
confidence levels.  Thus at $95\%$ confidence we rule out zero
dispersion only for the global sample and cannot state conclusively
that we resolve the central velocity dispersion.

Assuming the five central members are bound, we estimate the dynamical
mass of Leo V using a crude ``mass-follows-light'' (MFL) model in
which the density profile of any dark halo is proportional to that of
the stellar component, and which implies total mass $M=\eta
r_{h}\sigma_{V_0}^2$ \citep{illingworth76}.  For the purpose of
comparing to published estimates for other satellites we adopt
$\eta=850$M$_{\odot}$pc$^{-1}$km$^{-1}$s$^2$, the value corresponding
to a \citet{king62} profile with concentration parameter
characteristic of brighter
dSphs \citep{mateo98}.  With this value and the measured central velocity
dispersion we obtain $M=3.3_{-2.5(-3.3)}^{+9.1(+46)}\times
10^5M_{\odot}$ and $M_{\rm MFL}/L_V=75_{-58(-74)}^{+230(+1200)}$.  While
the lower mass limit should be interpreted as consistent with a purely
stellar population free of dark matter, even the upper limit gives Leo
V one of the lowest dynamical masses of the SDSS satellites with
measured kinematics (SG07, \citealt{martin07,geha08}).

\section{Metallicity}
\label{sec:metallicity}

\begin{figure}
\begin{center}
  \includegraphics[angle=270,scale=0.4]{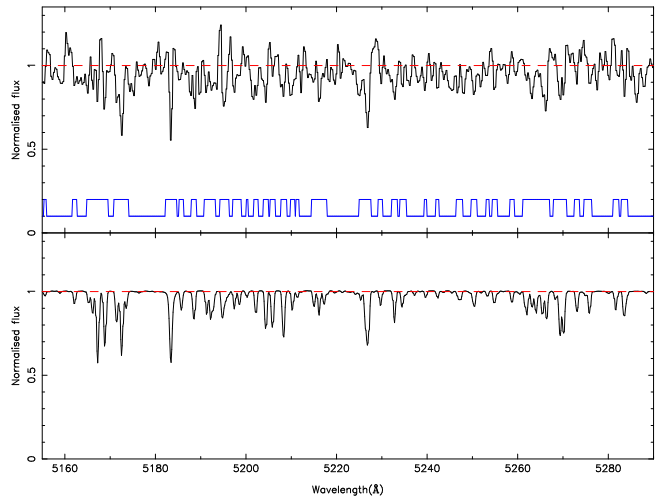}
\end{center}
  \caption{\textit{Top:} Stacked spectrum (continuum-normalised,
    averaged and rebinned) for the central 5 Leo V members.
    \textit{Bottom:} the best fit model spectrum; and in blue above
    the mask used in the minimisation.}
  \label{fig:feh}
\end{figure}

We estimate metallicity by comparing spectra to the library
of~\citet{lee08}, based on a regularly sampled grid on which
$[\alpha/\mathrm{Fe}] = 0.4$ covers a metallicity range $-4.0
<$[Fe/H]$< -0.5$, effective temperature $3500K < T_{\rm eff} < 9750$K,
and surface gravity $0.0 < \log g < 5.0$.  Since individual LeoV
spectra at $R=25000$ have insufficient S/N to perform a reliable
spectral analysis we averaged the 5 central member spectra and, after
suitable Gaussian smoothing, rebinned this average spectrum to a
resolution of $R=10000$.  After continuum normalizing both data and
model spectra, we compare the data directly to a suitable subset of
the synthetic spectra using a masked least-squares minimization.  The
adopted mask isolates regions where significant absorption lines
(typically with peaks $< 0.97$ of the continuum) appear in the model
spectrum (Figure \ref{fig:feh}).

This free-form fit gives a well-defined solution centered on $T_{\rm
  eff} = 5000\pm 250$K, [Fe/H]$= -2.0\pm 0.3$ dex and $\log g = 2.0\pm
1.0$.  Reassuringly, if we use the relationship $\log_{10}(T_{\rm
  eff}) = 3.877 - 0.26(g-r)$ \citep{ivezic06}, the average color
($\langle g-r\rangle =0.72$) of the five central members independently
implies $\langle T_{\rm eff}\rangle = 4900$K, consistent with the
temperature we derive directly from the continuum-normalized spectra.
Anchoring the value of $T_{\rm eff}$ for the spectral fit tightens
constraints on gravity and metallicity, giving $\log g =2.0 \pm 0.5$
and [Fe/H] $= -2.0 \pm 0.2$ dex.  The average color, magnitude,
surface gravity and effective temperature are fully consistent with
early K-giants at the distance of Leo V (see, e.g., \citealt{gray05},
p. 57).

\section{Discussion}
\label{sec:discussion}

Because we do not necessarily resolve the central velocity dispersion,
we can neither place a meaningful lower limit on the dynamical mass
nor state definitively that Leo V is a dark-matter-dominated dSph
rather than a diffuse star cluster devoid of dark matter.  In fact the
data lend some support to the latter interpretation, suggesting that
Leo V is an outlier with respect to two empirical scaling relations
among dSphs.  First, its dynamical mass-to-light ratio ($M/L_V=75_{-58}^{+233} [M/L_V]_{\odot}$) likely falls short of the more extreme
values ($\sim 10^{2-3}[M/L_V]_{\odot}$) exhibited by dSphs of
similar luminosity (e.g., Figure 15 of SG07).  Second, at
[Fe/H]$=-2.0\pm 0.2$ dex, Leo V is metal-rich compared to dSphs of
similar luminosity, which typically have [Fe/H]$\sim -2.5$ (SG07,
\citealt{kirby08}).

Despite ambiguities regarding the classification of Leo V, one
conclusion is clear: the available data are inconsistent with the
notion that Leo V is an equilibrium system with half-light radius
$r_{h}=0.8\arcmin \pm 0.1\arcmin$ ($\sim 50$ pc), the value measured
from the surface brightness of its red giants (Paper I).  Assuming Leo
V is a uniformly sampled (c.f. Figure \ref{fig:members_leo5}) Plummer
sphere, the probability of finding at least two members, in a sample
of seven, at $R\geq 10r_{\rm h}$ is $\sim 10^{-4}$.  Thus our detection of
two members at $R\sim 13\arcmin$ ($\sim 700$ pc) implies a distorted
morphology of the sort that might arise if the two outer members
belong to an unbound tidal stream.  The extreme velocities of these
two stars---they contribute the smallest and the largest velocities
among members---further suggests that they are unbound.

Tidal interactions with the MW produce stellar streams observable in
several MW satellites---for example, the Pal 5 globular cluster
\citep{odenkirchen03} and the Sagittarius dwarf galaxy (e.g.,
\citealt{majewski03}).  However, Leo V, at a distance of $D\sim 180$
kpc (Paper I), will be unaffected by MW tides unless its orbit is
strongly radial.  If Leo~V is on a radial orbit, then its journey from
near the center of the MW to its present position must have taken
$\sim 1$ Gyr. Assuming the two outer stars were once bound to Leo V
and were stripped at the most recent pericentric passage, then they
could have plausibly moved a distance of $\sim 700$ pc along the tidal
tail in this time (see Equation 7 of Odenkirchen et al. 2003).

We argue in Paper I that Leos IV and V are likely companions, in which
case their systemic velocities imply a nearly circular orbit that
would render MW tides unimportant.  In this case the two satellites
might interact with each other in filamentary substructure of the sort
produced in cosmological N-body simulations (e.g.,
\citealt{diemand07}).  The fact that Leo V's two outer stars lie along
the line connecting Leo IV to Leo V (Figure \ref{fig:members_leo5},
top) provides some support for this scenario, as do the extended
distributions of BHBs around both systems (Fig. 4 of Paper I).

Alternatively, Leo V may be a loosely bound star cluster losing stars
to evaporation.  It is not clear whether such a large cluster could
form in isolation, but Leo V may have been stripped from a progenitor
among the more luminous MW dSphs.  The best candidate is Leo II ($D
\sim 220$ kpc, $V_{\rm GSR} \sim 22$ km s$^{-1}$), which lies within $\sim
1.1^{\circ}$ of the orbit of the prospective Leo IV/V stream.  It is
also possible that a progenitor with low surface brightness lurks
outside the SDSS footprint, in which case it should be detectable with
data from upcoming deep-imaging surveys such as Pan-STARRS
\citep{kaiser02} and the Southern Sky Survey \citep{keller07}.

MGW acknowledges support from the STFC-funded Galaxy Formation and
Evolution programme at the Institute of Astronomy, Cambridge.  MM
acknowledges support from NSF grants AST-0206081 0507453, and 0808043.
EO acknowledges support from NSF Grants AST-0205790, 0505711, and
0807498.

\bibliography{ref}

\end{document}